\documentclass[a4paper,11pt]{article}
\usepackage{pos}
\usepackage{siunitx}
\usepackage{natbib}
\usepackage{graphicx}
\usepackage{subfig}
\usepackage{multirow}
\usepackage{booktabs}

\title{VERITAS Observations of the Microquasar SS 433}
 \ShortTitle{VERITAS Observations of the Microquasar SS 433}
\author*[a]{Tobias Kleiner}

\affiliation[a]{DESY,\\
  Platanenallee 6, 15738 Zeuthen, Germany}

\onbehalf{for the VERITAS Collaboration} 

\emailAdd{tobias.kleiner@desy.de}

\abstract{Microquasars are increasingly recognized as efficient particle accelerators, potentially contributing to the cosmic-ray flux up to the "knee". Among them, SS 433 stands out as a unique system with precessing relativistic jets embedded within the W50 supernova remnant. Recent detections of very- and ultra-high-energy gamma rays from SS 433 have solidified its role as a key laboratory for studying particle acceleration in jet-powered astrophysical sources.
We present results from over 100 hours of observations of SS 433 with VERITAS, spanning more than a decade. These high-resolution measurements allow for a detailed morphological study of the eastern and western jet lobes with an angular resolution of <0.1°. By analyzing the spatial and spectral characteristics of the gamma-ray emission, we investigate the particle acceleration mechanisms within the jets and at the jet-medium interaction in W50.}

\ConferenceLogo{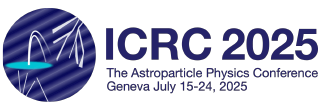}

\FullConference{39th International Cosmic Ray Conference (ICRC2025)\\
 15–24 July 2025\\
Geneva, Switzerland\\}

\begin{document}

\maketitle

\section{Introduction}

The VERITAS four-telescope imaging atmospheric Cherenkov telescope (IACT) array, located at the Fred Lawrence Whipple Observatory in southern Arizona (\SI{31}{\degree}40'N, \SI{110}{\degree} 57'W, 1.3 km a.s.l.), is sensitive to gamma rays in the energy range from \SI{100}{GeV} to >\SI{30}{TeV}, with an energy resolution of 15–\SI{25}{\percent} and an angular resolution of $<0.1^{\circ}$ (\SI{68}{\percent}) at \SI{1}{TeV}. VERITAS can detect a source with \SI{1}{\percent} of the Crab Nebula flux in \SI{25}{\hour}.\
The microquasar SS 433 is a remarkable astrophysical system for studying particle acceleration and transport in relativistic jets. Located at a distance of \SIrange{4.5}{5.5}{kpc} within the supernova remnant W50, it consists of a black hole accreting from an A-type donor star and ejecting precessing, baryon-loaded jets at initial speeds of $0.26\,c$. These jets propagate over tens of parsecs, interacting with the surrounding medium and producing emission across the electromagnetic spectrum \cite{marshallHighResolutionXRaySpectrum2002, margonEnormousPeriodicDoppler1979}.\
X-ray observations revealed distinct emission regions in both jet lobes and spectral hardening along the jet axis \cite{safi-harbHardXRayEmission2022}, with recent X-ray polarization measurements confirming synchrotron origin and magnetic fields aligned parallel to the jet flow \cite{kaaretXRayPolarizationEastern2024}. In the gamma-ray domain, SS 433 has emerged as a candidate Galactic PeVatron. HAWC reported TeV emission from both lobes \cite{abeysekaraVeryhighenergyParticleAcceleration2018}, and LHAASO-KM2A detected VHE and UHE gamma rays from both lobes and, more recently, also from the central source (above \SI{100}{TeV}) \cite{caoFirstLHAASOCatalog2023, lhaasocollaborationUltrahighEnergyGammarayEmission2024}, supporting evidence for PeV particle acceleration. H.E.S.S. observed extended TeV emission with energy-dependent morphology \cite{oliveranietoResolvingParticleAcceleration2023}, while Fermi-LAT detected GeV emission with indications of variability \cite{liGammarayHeartbeatPowered2020} with the precessional period of the jets.
These recent results highlight the role of microquasars in contributing to cosmic-ray flux up to the cosmic-ray knee and underscore the need for further high-resolution imaging atmospheric Cherenkov telescope observations.

\section{VERITAS Data Analysis}

\begin{figure}
	\includegraphics[width=0.55\linewidth]{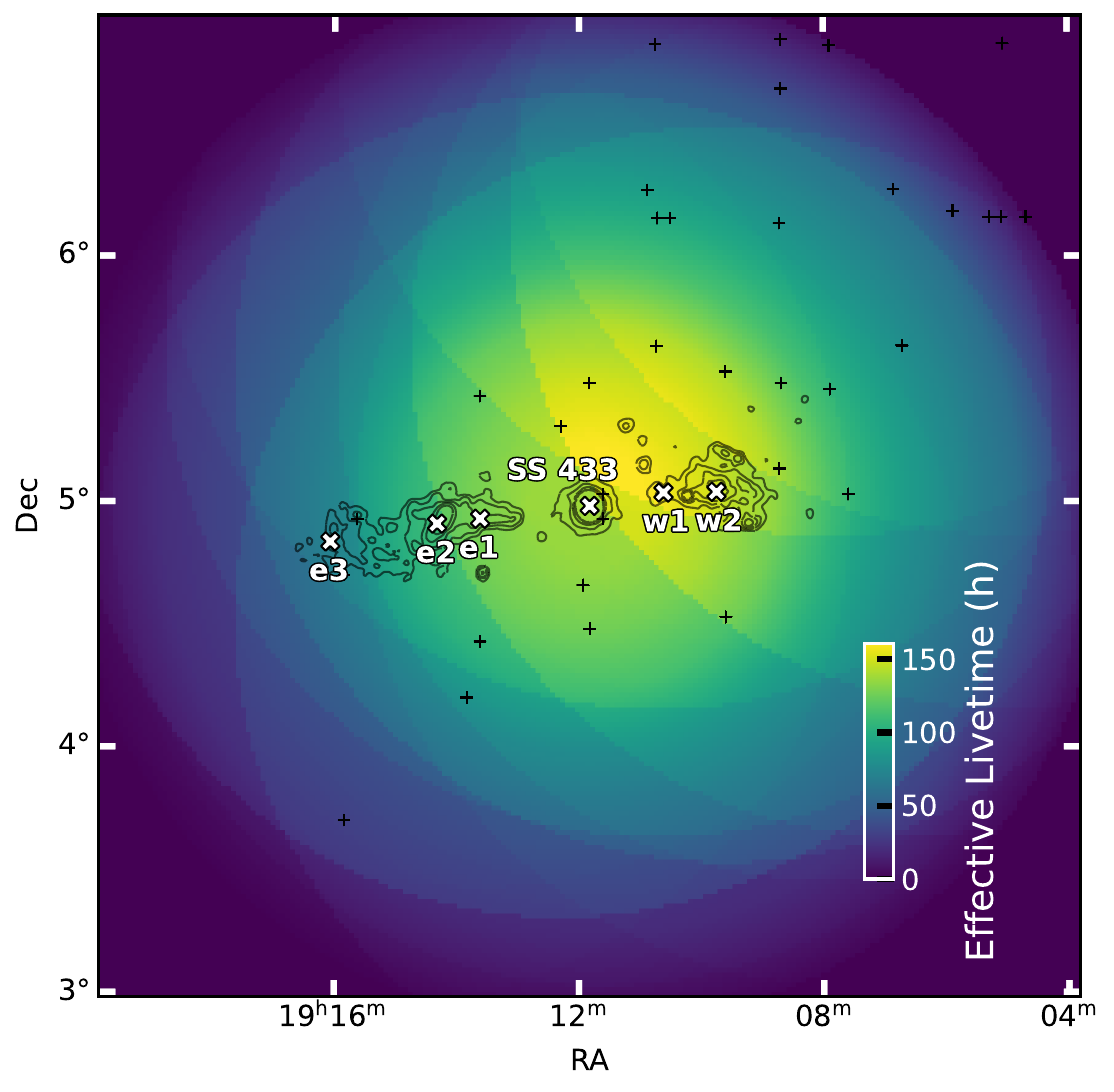}
	\centering
\caption[Acceptance corrected livetime]{SS 433 region acceptance corrected livetime map: Computed by dividing the exposure map by the VERITAS on axis effective area evaluated at an energy of \SI{1}{TeV}. The map is overlaid by black X-ray contours. The VERITAS observation pointings are indicated by black markers and SS 433, the eastern and western jet emission regions are indicated by white crosses.}
  	\label{fig:ss433:ss433_acceptance_corrected_exposure}
\end{figure}

VERITAS observed SS 433 over a period spanning from 2009 to September 2023. After correcting for data quality, weather, and instrument acceptance, the resulting exposures amount to approximately \SI{100}{\hour} for the eastern lobe and \SI{150}{\hour} for the western lobe, covering both outer jet regions (see Figure~\ref{fig:ss433:ss433_acceptance_corrected_exposure}).

Data are processed up to DL2 format using Eventdisplay (v491) \cite{maierEventdisplayAnalysisReconstruction2024}, which includes updates for improved sensitivity at $>1\,\mathrm{TeV}$, and converted to DL3 (V2DL3 v0.6.0). The final analysis, building on the analysis in \cite{kleinerInvestigatingMicroquasarSS2024} is performed with Gammapy (v1.3) using updated 3D maximum likelihood techniques in combination with a field-of-view background estimation method. The background models account for camera coordinates, energy, and observing conditions, including dependencies on zenith, azimuth, offset, and instrument epoch. Located less than a degree from SS 433, the strong and extended gamma-ray source MGRO J1908+06 \cite{acharyyaMultiwavelengthInvestigationGRay2024} must be accounted for in the analysis. It is modeled separately using a Gaussian spatial profile and a power-law spectral model and subtracted from the data to mitigate contamination.

Exclusion regions with radii of \SI{0.3}{\degree} are applied at the SS 433 central black hole and jet emission regions (e1–e3, w1–w2), and \SI{1.25}{\degree} around MGRO J1908+06. Additional regions are excluded for nearby pulsars.

\section{VERITAS Analysis Results}

\begin{figure}
  \centering
  \subfloat[]{
    \includegraphics[width=0.485\textwidth]{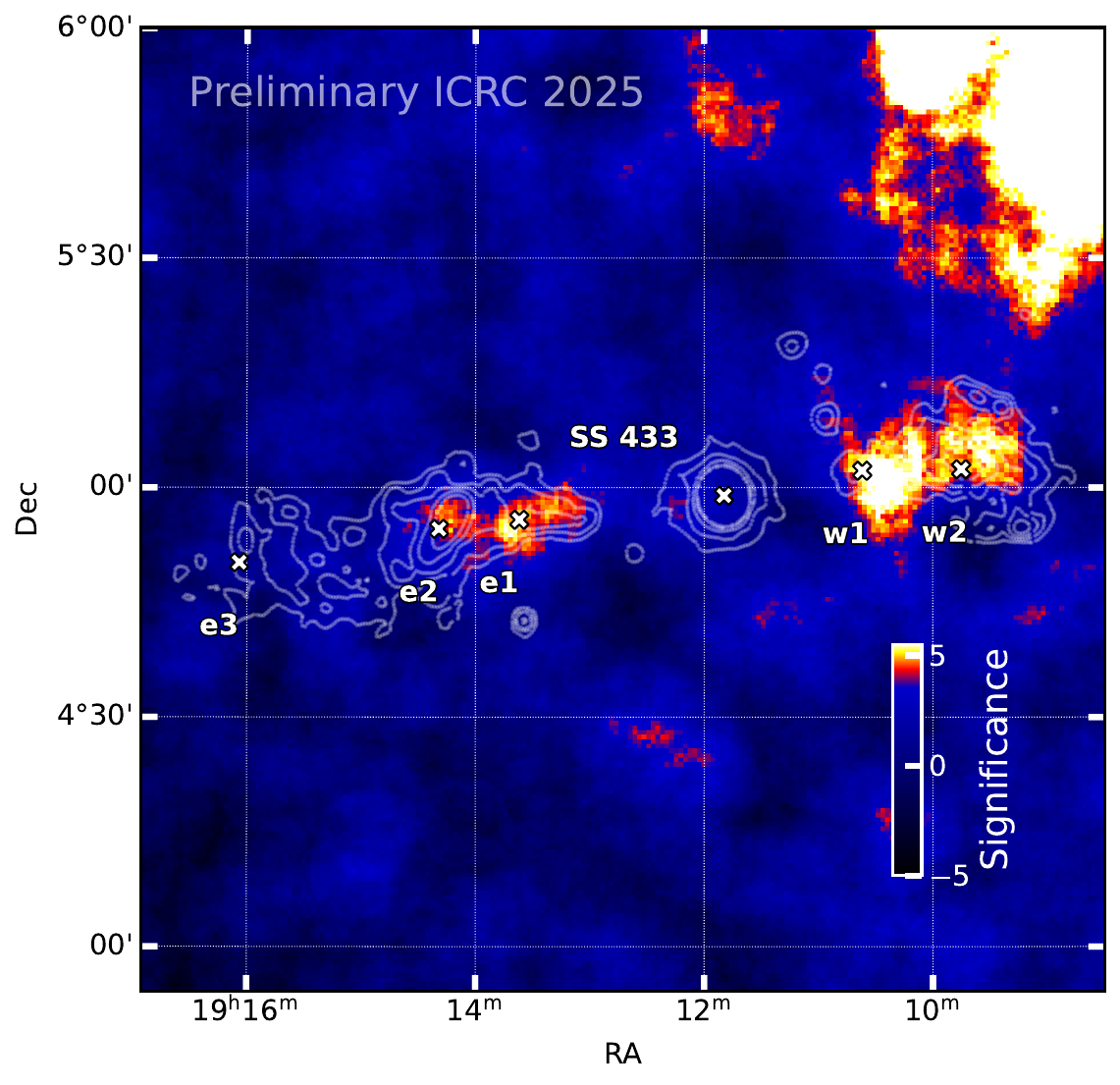}
  }
  \hfill
  \subfloat[]{
    \includegraphics[width=0.485\textwidth]{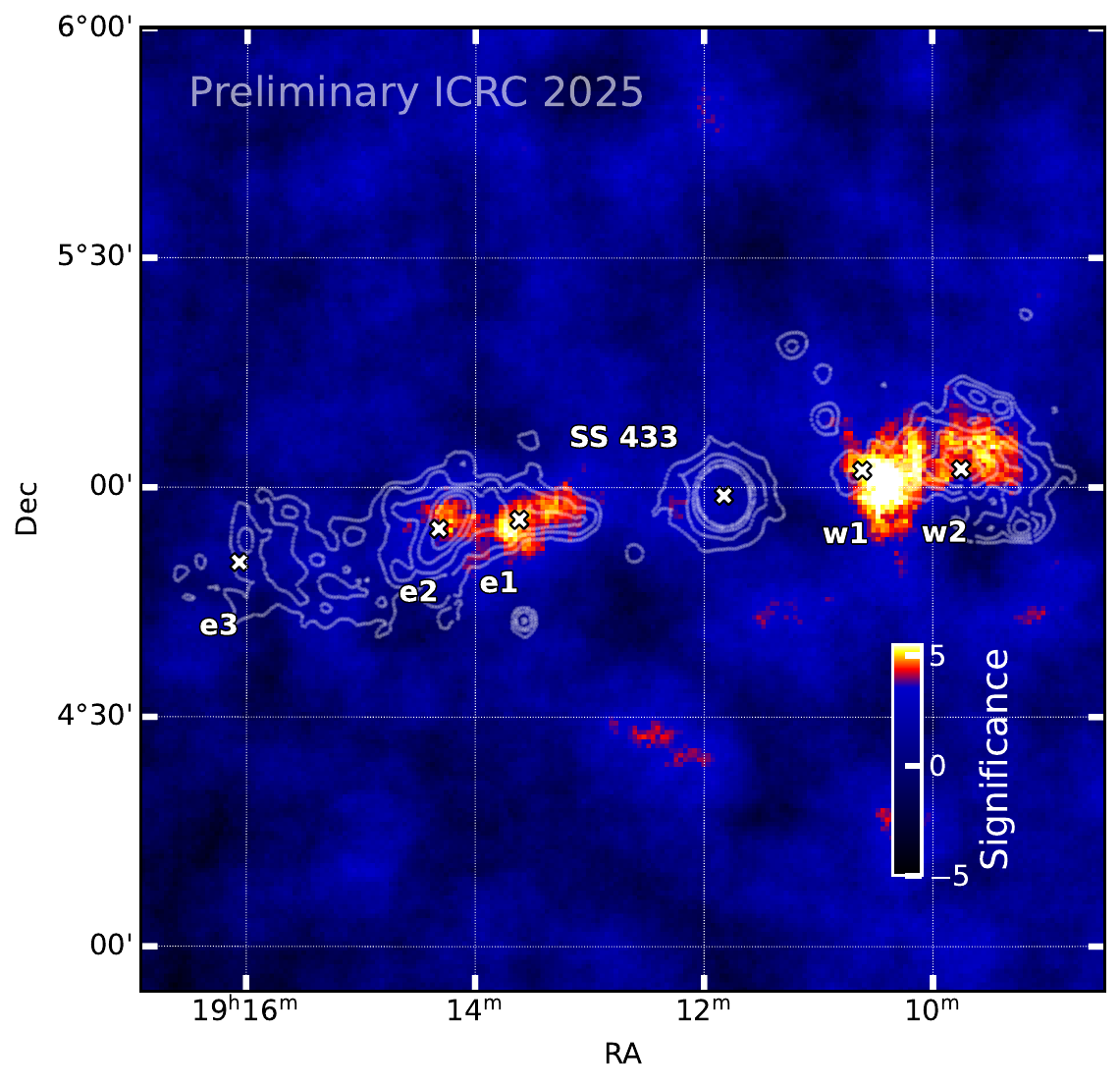}
  }
  \caption[VERITAS significance map of the microquasar SS~433]{VERITAS SS 433 significance maps before (a) and after (b) MGRO J1908+06 best-fit model subtraction. The eastern and western jet emission regions are marked by white crosses (from left to right: e3, e2, e1, w1, w2). White contours indicate ROSAT X-ray emission \cite{brinkmannROSATObservations501996}. Map bins have a size of \SI{0.01}{\degree} and are smoothed using a Top-Hat kernel with a radius of \SI{0.1}{\degree}.}
  \label{fig:ss433:ss433_sqrt_ts_map}
\end{figure}

The significance map (Figure~\ref{fig:ss433:ss433_sqrt_ts_map}), computed using the Cash statistic in the \SI{0.8}{TeV}–\SI{25}{TeV} range and smoothed with a Top-Hat kernel of radius \SI{0.1}{\degree}, reveals excess emission spatially coincident with the X-ray jet emission regions e1/e2 and w1/w2. After subtracting the MGRO J1908+06 best-fit model, significances of \SI{6.1}{\sigma} and \SI{5.5}{\sigma} are seen at the western and eastern jet lobes, respectively. No significant TeV emission is observed near the central black hole or beyond e2.

A joint spectro-morphological fit to the eastern and western lobes with an elongated Gaussian spatial model and a power-law spectral model yields a significance of $8.8\,\sigma$ over the null hypothesis. Individual fits show $6.0\,\sigma$ and $6.4\,\sigma$ significances for the eastern and western lobes, respectively. A symmetric extension is preferred over a point source at the $5.2\,\sigma$ (east) and $5.4\,\sigma$ (west) level. The elongated morphology is favored with $5.3\,\sigma$ jointly, and $4.4\,\sigma$ (east) and $3.6\,\sigma$ (west) individually.

\begin{table}[]
\centering
\begin{tabular}{ccccccc}
\hline
\textbf{Model}                & \textbf{R.A.} & \textbf{Dec.} & \textbf{Major axis}    & \textbf{Minor axis}    & \textbf{Eccentricity} & \textbf{Angle} \\ \hline
\multicolumn{1}{l}{\textbf{}} & \textbf{(deg)}           & \textbf{(deg)}       & \textbf{$1\,\boldsymbol{\sigma}$ (deg)} & \textbf{$1\,\boldsymbol{\sigma}$ (deg)} & \textbf{}             & \textbf{(deg)} \\ \hline
east            & $288.45\pm0.06$                   & $4.92\pm0.02$                 & $0.28\pm0.06$          & $0.06\pm0.02$              & $0.98\pm0.01$                  & $103.3\pm6.4$           \\
west            & $287.53\pm0.04$                   & $5.08\pm0.03$                 & $0.19\pm0.03$          & $0.11\pm0.02$             & $0.83\pm0.02$                  & $109.0\pm13.0$          \\ \hline
\end{tabular}
\caption[Elongated Gaussian model best fit parameters]{Best-fit parameters from a fit of the eastern and western jet emission regions of SS 433 with two elongated Gaussian functions.}
\label{tab:ss433_2gaussian_model_fit}
\end{table}

\begin{table}[h!]
\centering
\begin{tabular}{ccccc}
\hline
\textbf{Model}                & \textbf{Index} & \textbf{diff. Flux dn/dE} & \textbf{Flux $\geq 0.8\,\mathrm{TeV}$} & $\mathbf{E_{ref}}$ \\ \hline
\multicolumn{1}{l}{\textbf{}} & \textbf{}      & $\mathbf{10^{-14}{cm}^{-2}{s}^{-1}\mathrm{TeV}^{-1}}$      & $\mathbf{10^{-13}{cm}^{-2}{s}^{-1}}$                      & \textbf{(TeV)}     \\ \hline
SS 433 eastern jet            & $2.62\pm0.20$           & $1.48\pm0.37$                  & $4.53\pm1.60$                               & 4                  \\
SS 433 western jet            & $2.53\pm0.16$           & $1.55\pm0.28$                  & $4.21\pm1.23$                               & 4                  \\ \hline
\end{tabular}
\caption[Spectral parameters]{Spectral parameters obtained from a spectromorphological model, in which the spectrum is described by a power law and the morphology by a 2D elongated Gaussian distribution, fit to the eastern and western jet components of SS 433. The differential flux is given at the energy $\mathbf{E_{ref}}$.}
\label{tab:ss433_2gaussian_spectra}
\end{table}

\begin{figure}
	\includegraphics[width=1\linewidth]{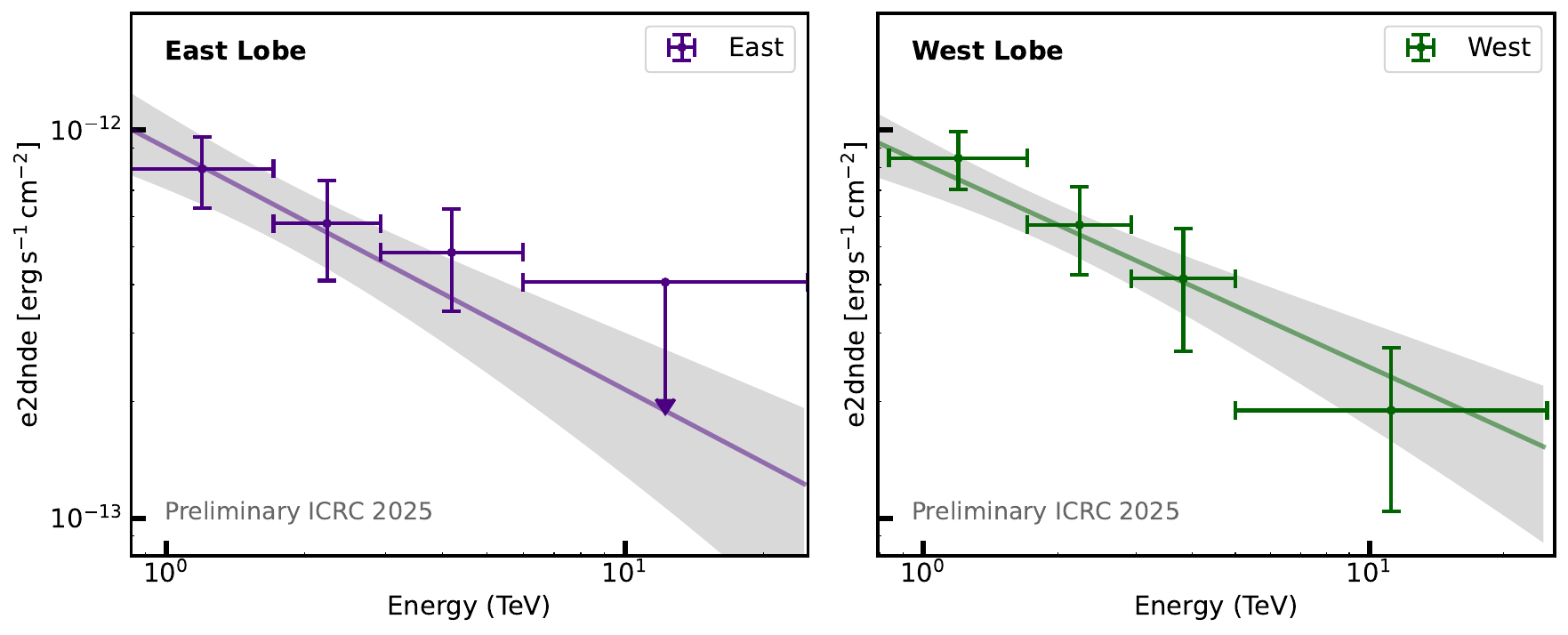}
	\centering
\caption[SS 433 differential flux points]{SS 433 differential flux points at the eastern and western outer jet lobes for the best-fit model.}
  	\label{fig:ss433:flux_comparison_east_west_jet_lobes}
\end{figure}

Flux points (see Figure~\ref{fig:ss433:flux_comparison_east_west_jet_lobes}) are extracted by fixing all parameters of the spectromorphological model, except for the power-law amplitude. The morphological parameters are taken from Table~\ref{tab:ss433_2gaussian_model_fit}, and the spectral indices are fixed as listed in Table~\ref{tab:ss433_2gaussian_spectra}.

To investigate potential variability, the dataset is split according to the orbital and precessional phases of the SS~433 system. For the eastern (e1e2) and western (w1w2) jet lobes, the same best-fit models are applied, and only the amplitude parameters are refitted in each phase bin. No evidence for flux variability is observed between the two precessional phase bins (0.0--0.5 and 0.5--1.0), which are selected according to \cite{liGammarayHeartbeatPowered2020}. 
A search for emission at the position of the central binary is performed using a point-like spatial- and a power-law spectral model with fixed indices of $\Gamma = 2$ and $\Gamma = 3$. No significant signal is detected in either precessional phase bin. The corresponding 95\% confidence level upper limits on the integral flux above 800\,GeV are $1.3 \times 10^{-13}\,\mathrm{cm}^{-2}\,\mathrm{s}^{-1}$ and $0.96 \times 10^{-13}\,\mathrm{cm}^{-2}\,\mathrm{s}^{-1}$ for phases 0.0--0.5, and $0.65 \times 10^{-13}\,\mathrm{cm}^{-2}\,\mathrm{s}^{-1}$ and $0.94 \times 10^{-13}\,\mathrm{cm}^{-2}\,\mathrm{s}^{-1}$ for phases 0.5--1.0, respectively.
A similar analysis is carried out using five orbital phase bins based on the $\sim$13-day orbital period. No significant flux variability is observed in these bins at either the jet lobes or the central region. The derived upper limits on the flux are $\leq 8.65 \times 10^{-14}\,\mathrm{cm}^{-2}\,\mathrm{s}^{-1}$.

\section{Broadband Modeling of SS~433's Jet Emission}

A multiwavelength spectral energy distribution (SED) for the eastern lobe of SS~433 is constructed using flux measurements from radio to very-high-energy (VHE) gamma rays, given the absence of significant spectral differences between the eastern and western jet lobes in VERITAS and the availability of data for the eastern lobe. The SED includes the derived VERITAS flux points, complemented by previously reported values from radio, X-ray, high-energy (HE), and VHE observations.
In the radio domain, a flux upper limit is adopted from \SI{2695}{MHz} observations with the Effelsberg radio telescope \citep{geldzahlerContinuumObservationsSNR1980}. For the soft X-ray regime ($2$--$10\,\mathrm{keV}$), XMM-Newton observations of the eastern jet region provide an unabsorbed flux of $7.5 \pm 0.2 \times 10^{-12}\,\mathrm{erg}\,\mathrm{cm}^{-2}\,\mathrm{s}^{-1}$ over the $0.5$--$10\,\mathrm{keV}$ range \citep{brinkmannXMMNewtonObservationsEastern2007, safi-harbHardXRayEmission2022}. In the hard X-ray band ($3$--$30\,\mathrm{keV}$), NuSTAR measurements of the full eastern jet region are included, as reported in \citep{safi-harbHardXRayEmission2022}.
The high-energy regime is represented by Fermi-LAT data, which reported a spectrum at the e1 position in a joint analysis with HAWC \citep{fangGeVTeVCounterparts2020}. In the VHE range, flux measurements are included from HAWC \citep{abeysekaraVeryhighenergyParticleAcceleration2018}, H.E.S.S. \citep{h.e.s.s.collaborationAccelerationTransportRelativistic2024}, and LHAASO \citep{lhaasocollaborationUltrahighEnergyGammarayEmission2024}. The VERITAS flux points are derived from the spectromorphological best-fit model in the $0.8$--$25\,\mathrm{TeV}$ energy range and include a systematic uncertainty of \SI{25}{\percent}.

\begin{figure}
	\includegraphics[width=0.6\linewidth]{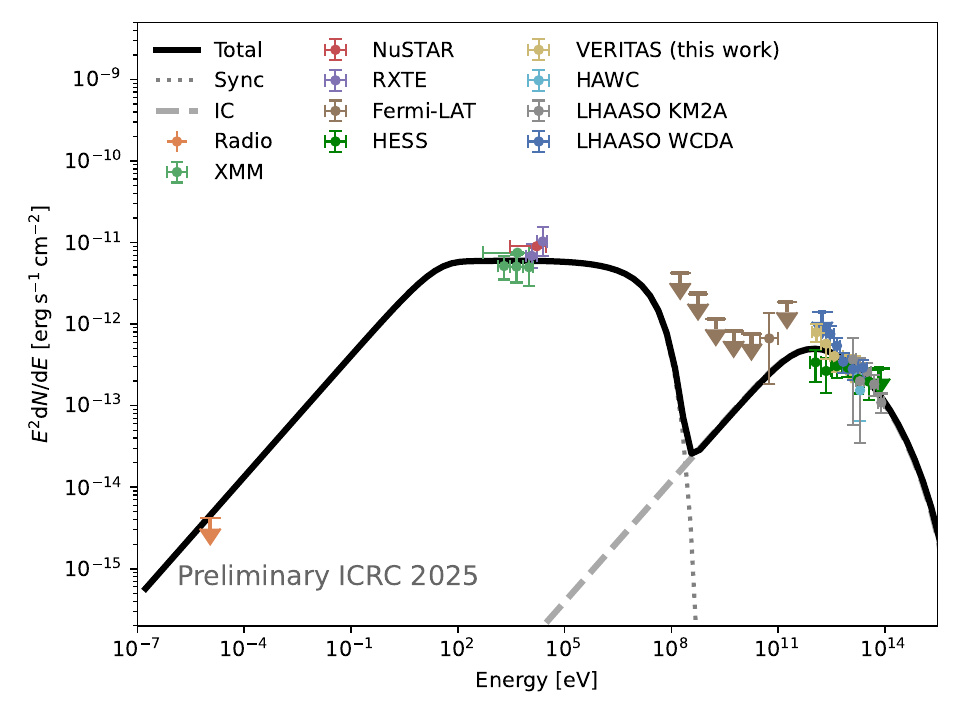}
	\centering
\caption[Leptonic SED with cooling effects]{The broadband spectral energy distribution model with cooling effects for the eastern jet emission region of SS 433 for an assumed source age of $10^4\,\mathrm{yr}$. The lines show the synchrotron and inverse Compton emission from the parent electron distribution.}
  	\label{fig:ss433:SS433_SynIC_gamera_evolved_pl_SED}
\end{figure}


A leptonic SED model is presented, in which electrons lose energy through synchrotron radiation, bremsstrahlung, ionization, and inverse Compton scattering. Electrons are injected following a power-law spectrum.
A static system is assumed, where the parameters of the power-law injection spectrum and the boundary conditions remain constant during the temporal evolution. CMB target photons are included with an energy density of $0.26\,\mathrm{eV}/\mathrm{cm}^3$. A distance of \SI{5.5}{kpc} to the source and an acceleration region size of \SI{6}{pc} is assumed. The time-evolved spectrum is computed for a continuous injection of electrons following a power-law distribution within this region.
Figure \ref{fig:ss433:SS433_SynIC_gamera_evolved_pl_SED} depicts the multiwavelength SED and the leptonic model. An energy break is visible at around \SI{1}{TeV} in this model.
The total energy in electrons amounts to $\sim3\times 10^{47}\,\mathrm{erg}$. 
This would involve the transfer of energy from the jets to the electron population at a sub-percent level to account for the non-thermal energy stored in the electron distribution of $\sim 10^{47}\,\mathrm{erg}$. This is a realistic scenario, taking into account the jets' kinetic power, which is estimated at $\sim 10^{50}-10^{51}\,\mathrm{erg}$ for an assumed source age between $10^4$ to $10^5$ years.
In contrast, using the derived proton density range of $n_{\mathrm{p}} = 0.15$--$1.2\,\mathrm{cm}^{-3}$ \citep{kleinerInvestigatingMicroquasarSS2024}, an extremely efficient conversion of the jet’s kinetic power into proton acceleration is required, potentially reaching or even exceeding the total kinetic energy output of the jets over the lifetime of the system.

\section{Conclusion and Discussion}

The detection of very-high-energy (VHE) gamma-ray emission from the microquasar SS 433 with VERITAS, at a combined statistical significance of \SI{8.8}{\sigma} from its eastern and western jet lobes, confirms that the jets of SS 433 accelerate particles to TeV energies. Using advanced 3D maximum likelihood techniques, the emission was shown to be best described by two elongated Gaussian morphologies aligned with the jet axis, indicating that the VHE emission originates from extended structures likely associated with jet-ISM interaction regions.
In the \SI{0.8}{TeV} to \SI{25}{TeV} energy range, no significant gamma-ray emission is detected from the central binary system, either in the full dataset or when split by precessional phase, and no evidence of variability with the orbital or precessional periods is observed. These findings indicate that the VHE emission arises from steady, large-scale acceleration processes at the jet termination regions, where the jets interact with the ambient medium. In the context of VERITAS observations, any periodic modulation from the central engine appears to be suppressed or smeared out during particle propagation to these outer jet regions, consistent with the lack of observed variability.

Multiwavelength modeling of the outer jet emission regions strongly favors a leptonic dominated emission scenario, placing constraints on hadronic contributions at these energies. The observed spatial extent and alignment, with the jets, provide compelling evidence for shock acceleration at jet termination sites interacting with the surrounding interstellar medium or from jet self-interaction.

These results firmly establish SS 433 as a benchmark system for VHE particle acceleration in microquasar jets and offer a key reference for jet-powered sources within the Galaxy. They underscore the importance of future high-sensitivity and high-resolution observations, such as with the Cherenkov Telescope Array Observatory (CTAO), to further resolve the emission morphology, constrain acceleration mechanisms, and explore possible neutrino counterparts in a multi-messenger context.

\section{Acknowledgements}

This research is supported by grants from the U.S. Department of Energy Office of Science, the U.S. National Science Foundation and the Smithsonian Institution, by NSERC in Canada, and by the Helmholtz Association in Germany. This research used resources provided by the Open Science Grid, which is supported by the National Science Foundation and the U.S. Department of Energy's Office of Science, and resources of the National Energy Research Scientific Computing Center (NERSC), a U.S. Department of Energy Office of Science User Facility operated under Contract No. DE-AC02-05CH11231. We acknowledge the excellent work of the technical support staff at the Fred Lawrence Whipple Observatory and at the collaborating institutions in the construction and operation of the instrument.

This research made use of gammapy,\footnote{https://www.gammapy.org} a community-developed core Python package for TeV gamma-ray astronomy \cite{deilGammapyPrototypeCTA2017}. This research made use of GAMERA \cite{hahnGAMERASourceModeling2022}. This research made use of the VERITAS analysis package Eventdisplay \cite{maierEventdisplayAnalysisReconstruction2024}.
\medskip

\bibliography{ss433_paper}

\clearpage
\section*{Full Author List: VERITAS Collaboration}

\scriptsize
\noindent
A.~Archer$^{1}$,
P.~Bangale$^{2}$,
J.~T.~Bartkoske$^{3}$,
W.~Benbow$^{4}$,
Y.~Chen$^{5}$,
J.~L.~Christiansen$^{6}$,
A.~J.~Chromey$^{4}$,
A.~Duerr$^{3}$,
M.~Errando$^{7}$,
M.~Escobar~Godoy$^{8}$,
J.~Escudero Pedrosa$^{4}$,
Q.~Feng$^{3}$,
S.~Filbert$^{3}$,
L.~Fortson$^{9}$,
A.~Furniss$^{8}$,
W.~Hanlon$^{4}$,
O.~Hervet$^{8}$,
C.~E.~Hinrichs$^{4,10}$,
J.~Holder$^{11}$,
T.~B.~Humensky$^{12,13}$,
M.~Iskakova$^{7}$,
W.~Jin$^{5}$,
M.~N.~Johnson$^{8}$,
E.~Joshi$^{14}$,
M.~Kertzman$^{1}$,
M.~Kherlakian$^{15}$,
D.~Kieda$^{3}$,
T.~K.~Kleiner$^{14}$,
N.~Korzoun$^{11}$,
S.~Kumar$^{12}$,
M.~J.~Lang$^{16}$,
M.~Lundy$^{17}$,
G.~Maier$^{14}$,
C.~E~McGrath$^{18}$,
P.~Moriarty$^{16}$,
R.~Mukherjee$^{19}$,
W.~Ning$^{5}$,
R.~A.~Ong$^{5}$,
A.~Pandey$^{3}$,
M.~Pohl$^{20,14}$,
E.~Pueschel$^{15}$,
J.~Quinn$^{18}$,
P.~L.~Rabinowitz$^{7}$,
K.~Ragan$^{17}$,
P.~T.~Reynolds$^{21}$,
D.~Ribeiro$^{9}$,
E.~Roache$^{4}$,
I.~Sadeh$^{14}$,
L.~Saha$^{4}$,
H.~Salzmann$^{8}$,
M.~Santander$^{22}$,
G.~H.~Sembroski$^{23}$,
B.~Shen$^{12}$,
M.~Splettstoesser$^{8}$,
A.~K.~Talluri$^{9}$,
S.~Tandon$^{19}$,
J.~V.~Tucci$^{24}$,
J.~Valverde$^{25,13}$,
V.~V.~Vassiliev$^{5}$,
D.~A.~Williams$^{8}$,
S.~L.~Wong$^{17}$,
T.~Yoshikoshi$^{26}$\\
\\
\noindent
$^{1}${Department of Physics and Astronomy, DePauw University, Greencastle, IN 46135-0037, USA}

\noindent
$^{2}${Department of Physics, Temple University, Philadelphia, PA 19122, USA}

\noindent
$^{3}${Department of Physics and Astronomy, University of Utah, Salt Lake City, UT 84112, USA}

\noindent
$^{4}${Center for Astrophysics $|$ Harvard \& Smithsonian, Cambridge, MA 02138, USA}

\noindent
$^{5}${Department of Physics and Astronomy, University of California, Los Angeles, CA 90095, USA}

\noindent
$^{6}${Physics Department, California Polytechnic State University, San Luis Obispo, CA 94307, USA}

\noindent
$^{7}${Department of Physics, Washington University, St. Louis, MO 63130, USA}

\noindent
$^{8}${Santa Cruz Institute for Particle Physics and Department of Physics, University of California, Santa Cruz, CA 95064, USA}

\noindent
$^{9}${School of Physics and Astronomy, University of Minnesota, Minneapolis, MN 55455, USA}

\noindent
$^{10}${Department of Physics and Astronomy, Dartmouth College, 6127 Wilder Laboratory, Hanover, NH 03755 USA}

\noindent
$^{11}${Department of Physics and Astronomy and the Bartol Research Institute, University of Delaware, Newark, DE 19716, USA}

\noindent
$^{12}${Department of Physics, University of Maryland, College Park, MD, USA }

\noindent
$^{13}${NASA GSFC, Greenbelt, MD 20771, USA}

\noindent
$^{14}${DESY, Platanenallee 6, 15738 Zeuthen, Germany}

\noindent
$^{15}${Fakult\"at f\"ur Physik \& Astronomie, Ruhr-Universit\"at Bochum, D-44780 Bochum, Germany}

\noindent
$^{16}${School of Natural Sciences, University of Galway, University Road, Galway, H91 TK33, Ireland}

\noindent
$^{17}${Physics Department, McGill University, Montreal, QC H3A 2T8, Canada}

\noindent
$^{18}${School of Physics, University College Dublin, Belfield, Dublin 4, Ireland}

\noindent
$^{19}${Department of Physics and Astronomy, Barnard College, Columbia University, NY 10027, USA}

\noindent
$^{20}${Institute of Physics and Astronomy, University of Potsdam, 14476 Potsdam-Golm, Germany}

\noindent
$^{21}${Department of Physical Sciences, Munster Technological University, Bishopstown, Cork, T12 P928, Ireland}

\noindent
$^{22}${Department of Physics and Astronomy, University of Alabama, Tuscaloosa, AL 35487, USA}

\noindent
$^{23}${Department of Physics and Astronomy, Purdue University, West Lafayette, IN 47907, USA}

\noindent
$^{24}${Department of Physics, Indiana University Indianapolis, Indianapolis, Indiana 46202, USA}

\noindent
$^{25}${Department of Physics, University of Maryland, Baltimore County, Baltimore MD 21250, USA}

\noindent
$^{26}${Institute for Cosmic Ray Research, University of Tokyo, 5-1-5, Kashiwa-no-ha, Kashiwa, Chiba 277-8582, Japan}

%
%
%

\end{document}